\documentclass[a4paper,aps,prl,superscriptaddress,onecolumn]{revtex4}

\usepackage{graphicx} 
\usepackage[usenames,dvipsnames]{xcolor}
\usepackage{tikz}
\usepackage{amsmath}
\usepackage{tabularx}
\usepackage{amssymb}
\usepackage{natbib}
\usepackage{color}
\usepackage{psfrag}
\usepackage{wasysym}
\usepackage{float}
\usepackage{epstopdf}
\usepackage[normalem]{ulem}
\usepackage{array}
\usepackage{wrapfig}
\usepackage{comment}
\usepackage{siunitx}

\newcolumntype{+}{>{\global \let \currentrowstyle \relax}}
\newcolumntype{^}{>{\currentrowstyle}}

\begin{document}
\title{Equilibria and Instabilities of a Slinky: Discrete Model}
\author{Douglas P. Holmes}
\affiliation{Department of Engineering Science and Mechanics, Virginia Tech, Blacksburg, VA 24061, USA}
\author{Andy D. Borum}
\affiliation{Department of Aerospace Engineering, University of Illinois at Urbana-Champaign, Urbana, IL 61801, USA}
\author{Billy F. Moore III}
\affiliation{Department of Engineering Science and Mechanics, Virginia Tech, Blacksburg, VA 24061, USA}
\author{Raymond H. Plaut}
\affiliation{Department of Civil and Environmental Engineering, Virginia Tech, Blacksburg, VA 24061, USA}
\author{David A. Dillard}
\affiliation{Department of Engineering Science and Mechanics, Virginia Tech, Blacksburg, VA 24061, USA}

\date{\today}

\begin{abstract}
The Slinky is a well-known example of a highly flexible helical spring, exhibiting large, geometrically nonlinear deformations from minimal applied forces. By considering it as a system of coils that act to resist axial, shearing, and rotational deformations, we develop a discretized model to predict the equilibrium configurations of a Slinky via the minimization of its potential energy. Careful consideration of the contact between coils enables this procedure to accurately describe the shape and stability of the Slinky under different modes of deformation. In addition, we provide simple geometric and material relations that describe a scaling of the general behavior of flexible, helical springs.
\end{abstract}

\pacs{}

\maketitle

\begin{figure}
\begin{center}
\resizebox{1\textwidth}{!} {\includegraphics {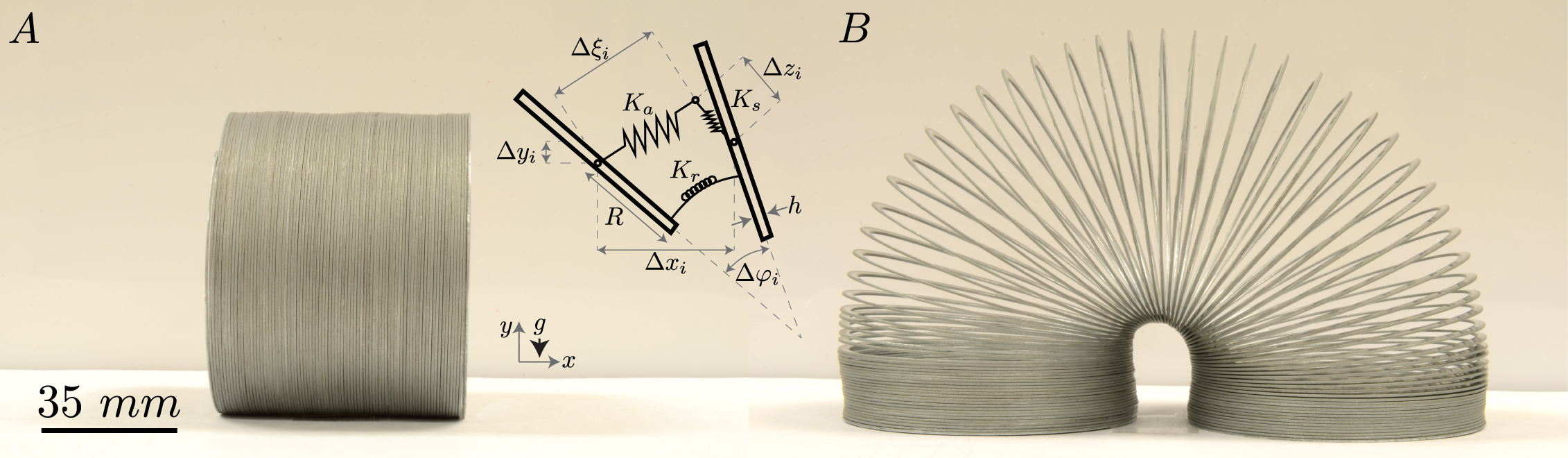}}
\end{center}
\vspace{-6mm}
\caption[]{({\bf A}) A Slinky on a flat surface in two stable states, and ({\bf B}) an accompanying schematic showing bar $i$ (on left) and bar $i+1$ (on right) for the discrete model, along with displacements and axial, rotational, and shear springs. \vspace{0mm}
\label{fig1}}
\end{figure}

The floppy nature of a tumbling Slinky (Poof-Slinky, Inc.) has captivated children and adults alike for over half a century. Highly flexible, the spring will walk down stairs, turn over in your hands, and -- much to the chagrin of children everywhere - become easily entangled and permanently deformed. The Slinky can be used as an educational tool for demonstrating standing waves, and a structural inspiration due to its ability to extend many times beyond its initial length without imparting plastic strain on the material.  Engineers have scaled the iconic spring up to the macroscale as a pedestrian bridge~\cite{Schlaich2012}, and down to the nanoscale for use as conducting wires within flexible electronic devices~\cite{Sun2006, Xu2011}, while animators have simulated its movements in a major motion picture~\cite{Lo2010}.  Yet, perhaps the most recognizable and remarkable features of a Slinky are simply its ability to splay its helical coils into an arch (Fig.~\ref{fig1}), and to tumble over itself down a steep incline. 

A 1947 patent by Richard T. James for ``Toy and process of use"~\cite{James1947} describes what became known as the Slinky, ``a helical spring toy adapted to walk and oscillate." The patent discusses the geometrical features, such as a rectangular cross section with a width-to-thickness ratio of 4:1, compressed height approximately equal to the diameter, almost no pretensioning but adjacent turns (coils) that touch each other in the absence of external forces, and the ability to remain in an arch shape on a horizontal surface. In the same year, Cunningham~\cite{Cunningham1947} performed some tests and analysis of a steel Slinky tumbling down steps and down an inclined plane. His steel Slinky had 78 turns, a length of 6.3 cm, and an outside diameter of 7.3 cm. He examined the spring stiffness, the effects of different step heights and of inclinations of the plane, the time length per tumble and the corresponding angular velocity, and the velocity of longitudinal waves. He stated that the time period for a step height between 5 and 10 cm is almost independent of the height and is about 0.5 s. Forty years later, he gave a further description of waves in a tumbling Slinky~\cite{Cunningham1987}. Longuet-Higgins~\cite{LonguetHiggins1954} also studied a Slinky tumbling down stairs. His phosphor-bronze Slinky had 89 turns, a length of 7.6 cm, and an outside diameter of 6.4 cm. In his analysis, he imagined the Slinky as an elastic fluid, with one density at the end regions where coils touch and another for the rest. His tests produced an average time of about 0.8 s per step for a variety of step heights.

Heard and Newby~\cite{Heard1977} hung a Slinky-like spring vertically, held at its top, with and without a mass attached at the bottom. Using experiments and analysis, they investigated the length, as did French~\cite{French1994}, Sawicki~\cite{Sawicki2002}, and Gluck~\cite{Gluck2010}, and they studied longitudinal waves, as did Young~\cite{Young1993}, Bowen~\cite{Bowen1982}, and Gluck~\cite{Gluck2010}. In the work by Bowen, the method of characteristics was utilized to obtain solutions of the wave equation (see also~\cite{Cushing1984}), and an effective mass of the Slinky was discussed, which was related to the weight applied to an associated massless spring and yielding the same fundamental vibration period. Mak~\cite{Mak1987} defined an effective mass with regard to the static elongation of the vertically suspended Slinky. Blake and Smith~\cite{Blake1979} and Vandergrift \textit{et al.}~\cite{Vandegrift1989} suspended a Slinky horizontally by strings and investigated the behavior of transverse vibrations and waves. Longitudinal and transverse waves in a horizontal Slinky were examined by Gluck~\cite{Gluck2010}. Crawford~\cite{Crawford1987} discussed ``whistler" sounds produced by longitudinal and transverse vibrations of a Slinky held at both ends. Musical sounds that could be obtained from a Slinky were described by Parker \textit{et al.}~\cite{Parker2010}, and Luke~\cite{Luke1992} considered a Slinky-like spring held at its ends in a U shape and the propagation of pulses along the spring. Wilson~\cite{Wilson2001} investigated the Slinky in its arch configuration. In his analysis, each coil was modeled as a rectangular bar, and a rotational spring connected each pair of adjacent bars. Some bars at the bottom of each end (leg) of the arch were in full horizontal contact with each other due to the pretensioning of the spring. The angular positions of the bars were computed for springs with 87 and 119 coils, and were compared with experimental results. Wilson also lowered one end quasi-statically until the Slinky tumbled over that end. The discrete model in the present paper will be an extension of Wilson's model and will include rotational, axial, and shear springs connecting adjacent bars.


Hu~\cite{Hu2010} analyzed a simple two-link, two-degree-of-freedom model of a Slinky walking down stairs. The model included a rotational spring and rotational dashpot at the hinge that connected the massless rigid links, with equal point masses at the hinge and the other end of each link. The equations of motion for the angular coordinates of the bars were solved numerically. Periodic motion was predicted for a particular set of initial conditions.  The apparent levitation of the Slinky's bottom coils as the extended spring is dropped in a gravitational field has proved both awe-inspiring and confounding~\cite{Calkin1993, Gardner2000, Graham2001, Kolkowitz2007, Aguirregabiria2007, Unruh2011, Cross2012, Sakaguchi2013, Plaut2014}. If a Slinky is held at its top in a vertical configuration and then released, it has been shown that its bottom does not move for a short amount of time as the top part drops. A slow-motion video has been used to demonstrate this phenomenon~\cite{Shomsky2011}. 

A Slinky is a soft, helical spring made with wire of rectangular cross section. The mechanics of helical springs has been studied since the time of Kirchhoff~\cite{Todhunter1893}, and their nonlinear deformations were first examined in the context of elastic stability.  The spring's elastic response to axial and transverse loading was first characterized by treating it as a prismatic rod and ignoring the transverse shear elasticity of the spring~\cite{Hurlbrink1910,Grammel1924}. Modifications to these equilibrium equations initially over-estimated the importance of shear~\cite{Biezeno1925}, thereby implying that buckling would occur for any spring, regardless of its length. The contribution from a spring's shear stiffness was properly accounted for by Haringx~\cite{Haringx1948} and Ziegler and Huber~\cite{Ziegler1950}, which enabled an accurate prediction of the elastic stability of highly compressible helical springs. Large, nonlinear deformations of stiff springs occur when lateral buckling thresholds are exceeded in tension~\cite{Kessler2003} or compression~\cite{Haringx1948, Wahl1963}. Soft helical springs, with a minimal resistance to axial and bending deformations, may exhibit large deformations from the application of very little force. It can be readily observed with a Slinky that small changes in applied load can lead to significant nonlinear deformations. Simplified energetic models have been developed to capture the nonlinear deformations of soft helical springs~\cite{Wilson2001}.

Recent experimental work has focused on fabricating and characterizing helical springs on the nanoscale. Their potential usefulness in nanoelectromechanical systems (NEMS) as sensors and actuators has led to extensive developments in recent years~\cite{Korgel2005} using carbon~\cite{Poggi2004}, zinc oxide~\cite{Gao2005}, Si/SiGe bilayers~\cite{Kim2011a}, and CdSe quantum dots~\cite{Pham2013} to form nanosprings. The mechanical properties of these nanosprings, including the influence of surface effects on spring stiffness~\cite{Wang2010, Wang2011}, has been evaluated at an atomistic level~\cite{Chang2008}, as amorphous structures~\cite{Da2006}, and as viscosity modifiers within polymeric systems~\cite{Liu2012}. Recently, nanosprings or nanoparticle helices were fabricated by utilizing a geometric asymmetry, and were shown to be highly deformable, soft springs~\cite{Pham2013}.

In this paper, we provide a mechanical model that captures the static equilibrium configurations of the Slinky in terms of its geometric and material properties. In section I, we consider a discretized model in which the Slinky is represented as a series of rigid bars connected by springs that resist axial, shear, and rotational deformations. In section II, we provide a means for determining the effective spring stiffnesses based on three static equilibrium shapes. Finally, in section III, we compare experimental results obtained for the Slinky's static equilibrium shapes, and we determine the critical criteria for the Slinky to topple over in terms of the vertical displacement of one base of the arch, and the critical number of cantilevered coils.

\section{I. Discrete Model}

In order to adequately account for the contact between Slinky coils, and the effect this contact has on the Slinky's equilibrium shapes, we introduce a discretized model that represents an extension of Wilson's model~\cite{Wilson2001}.  The total effective energy $V$ of a Slinky is comprised of its elastic and gravitational potential energies. Friction between individual coils, and along the contact surface, further complicates this energetic analysis, and is neglected in our calculations. In this discretized model, the coils are represented by rigid bars, with the centers of adjacent bars connected by axial, rotational, and shear springs.  Each translational spring is assumed to be unstretched when its length is zero, and each rotational spring is assumed to be unstretched when its angle of splay is zero. The elastic energy of a Slinky with $n$ coils is the sum of the strain energy associated with axial, rotational, and shear deformations (Fig.~\ref{fig1}B), with stiffnesses denoted by $K_a$, $K_r$, and $K_s$, respectively. We can separate the displacement between two adjacent bars into individual components that correspond to deformations of \textit{effective} axial, rotational, and shear springs that connect each coil. We denote $\Delta \xi_i$ as the extension of the axial spring, $\Delta \varphi_i$ as the difference between the angles of the bars connected to the rotational spring, and $\Delta z_i$ as the extension of the shear spring occurring between adjacent bars (Fig.~\ref{fig1}B). The axial and shear deformations can be determined from a geometric relationship by
\begin{subequations}
\begin{align}
\label{eq-geo}
\Delta \xi_i&=\Delta x_i \sin \left (\frac{\varphi_{i+1}+\varphi_i}{2}\right)+\Delta y_i \cos \left (\frac{\varphi_{i+1}+\varphi_i}{2}\right), \\ 
\Delta z_i&=\Delta x_i \cos \left (\frac{\varphi_{i+1}+\varphi_i}{2}\right)-\Delta y_i \sin \left (\frac{\varphi_{i+1}+\varphi_i}{2}\right), 
\end{align}
\end{subequations}
where $\Delta x_i$ and $\Delta y_i$ in Fig.~\ref{fig1} are the differences in horizontal and vertical coordinates, respectively, between the centers of mass of bars $i$ and $i+1$, and $\varphi_i$ is the angle between the $-x$ axis and bar $i$, positive if clockwise. 

Boundary conditions can be prescribed on the variables $x_i$, $y_i$, or $\varphi_i$ for some of the bars. For instance, for the splayed Slinky in Fig.~\ref{fig1}, the boundary conditions at the left end would be $x_1=y_1=\varphi_1=0$, and at the right end they would be $y_n = 0$, $\varphi_n=180^o$, and $x_n = 2R+c_0$, where $R$ is the radius of the Slinky (and half the length of each bar), and $c_0$ is some positive constant. 

Equilibrium shapes of this system of springs and masses can be found by minimizing $V$ with respect to all unprescribed variables. The effective potential energy, including the gravitational potential energy, is written as
\begin{equation}
\label{eq-V}
V=\frac{1}{2}K_a\sum_{i=1}^{n-1}\left(\Delta \xi_i+\frac{mgn_p}{K_a}\right)^2+\frac{1}{2}K_s\sum_{i=1}^{n-1}\Delta z_i^2+\frac{1}{2}K_r\sum_{i=1}^{n-1}\Delta \varphi_i^2+mg\sum_{i=1}^{n}y_i,
\end{equation}
where $m$ is the mass per coil, and $g$ is the acceleration in the $-y$ direction due to gravity. We assume that pretensioning of the Slinky causes a constant precompression force $P_p$ and, when the Slinky hangs vertically, causes $n_p$ coils at the bottom to be compressed together~\cite{Mak1987}. The precompression force is approximately equal to the weight of these compressed coils, $i.e.$, $P_p = mgn_p$. The axial term in equation~\ref{eq-V} includes the deformation required to overcome $P_p$.

Accounting for the elastic potential energy of the springs alone will only correspond to equilibrium shapes in the regime where there is no contact between Slinky coils. The contact between coils adds a nonlinearity that is not accounted for in equation~\ref{eq-V}.  Two types of contact can occur along the extended length of the spring. The first type, which we refer to as axial contact, occurs when two adjacent coils are in contact around the entire circumference of the Slinky, as seen in the legs of the arch in Fig.~\ref{fig1}. The second type, which we refer to as rotational contact, occurs when two adjacent coils touch at only one point along the circumference, as seen in the coils above the legs of the arch in Fig.~\ref{fig1}. To enforce the axial contact constraint, we must ensure that the axial deformation is never smaller than the thickness, $i.e.$  $\Delta \xi_i \geq h$. This is done by introducing a penalty function of the form
\begin{equation}
P_a=\alpha_a\sum_{i=1}^{n-1}\max\left(0,-\left(\Delta \xi_i-h\right)\right)^2,
\end{equation}
where $\alpha_a$ controls the weight of the axial contact penalty function. To account for rotational contact, consider Fig.~\ref{fig1}B with the lower end of the left bar in contact with the right bar. In this configuration, $\Delta \xi_i=\Delta \xi_{\min}$ where
\begin{equation}
\label{eq-xi}
\Delta \xi_{\min} = 2R\sin \frac{\Delta \varphi_i}{2}+h\cos \frac{\Delta \varphi_i}{2}-\Delta z_i \tan \frac{\Delta \varphi_i}{2}.
\end{equation}
Therefore, for a given $\Delta z_i$ and $\Delta \varphi_i$, $\Delta \xi_{\min}$ is the minimum admissible axial deformation. We can impose the constraint that $\Delta \xi_i > \Delta \xi_{\min}(\pm \Delta \varphi_i, \Delta z_i)$ with the penalty function 
\begin{equation}
\label{eq-Pr}
P_r=\alpha_r\sum_{i=1}^{n-1}\max\left(0,-\left(\Delta \xi_i-\Delta \xi_{\min}\left(\pm \Delta \varphi_i, \Delta z_i\right)\right)\right)^2, 
\end{equation}
where $\alpha_r$ controls the weight of the rotational contact penalty function. These two additional energetic penalties enable us to define the augmented total potential energy $E$ as
\begin{equation}
\label{eq-E}
E=V+P_a+P_r
\end{equation}
The local minima of $E$ with respect to all unprescribed $x_i$, $y_i$, and $\varphi_i$ yield predictions for stable equilibrium shapes of the Slinky.

\section{II. Spring Stiffnesses and Equilibria}

The augmented total potential energy is dependent on the stiffnesses of the springs. We will determine the relevant spring stiffnesses based on simple mechanical equilibrium of the Slinky structure in three specific configurations. The benefit of the static equilibria method is its ease of implementation for flexible springs large enough to have gravity be the dominant body force, while single coil analysis via Castigliano's method provides a scalable means for determining the relevant spring stiffnesses~\cite{Borum2014}.

\begin{figure}[b]
\begin{center}
\resizebox{1\textwidth}{!} {\includegraphics {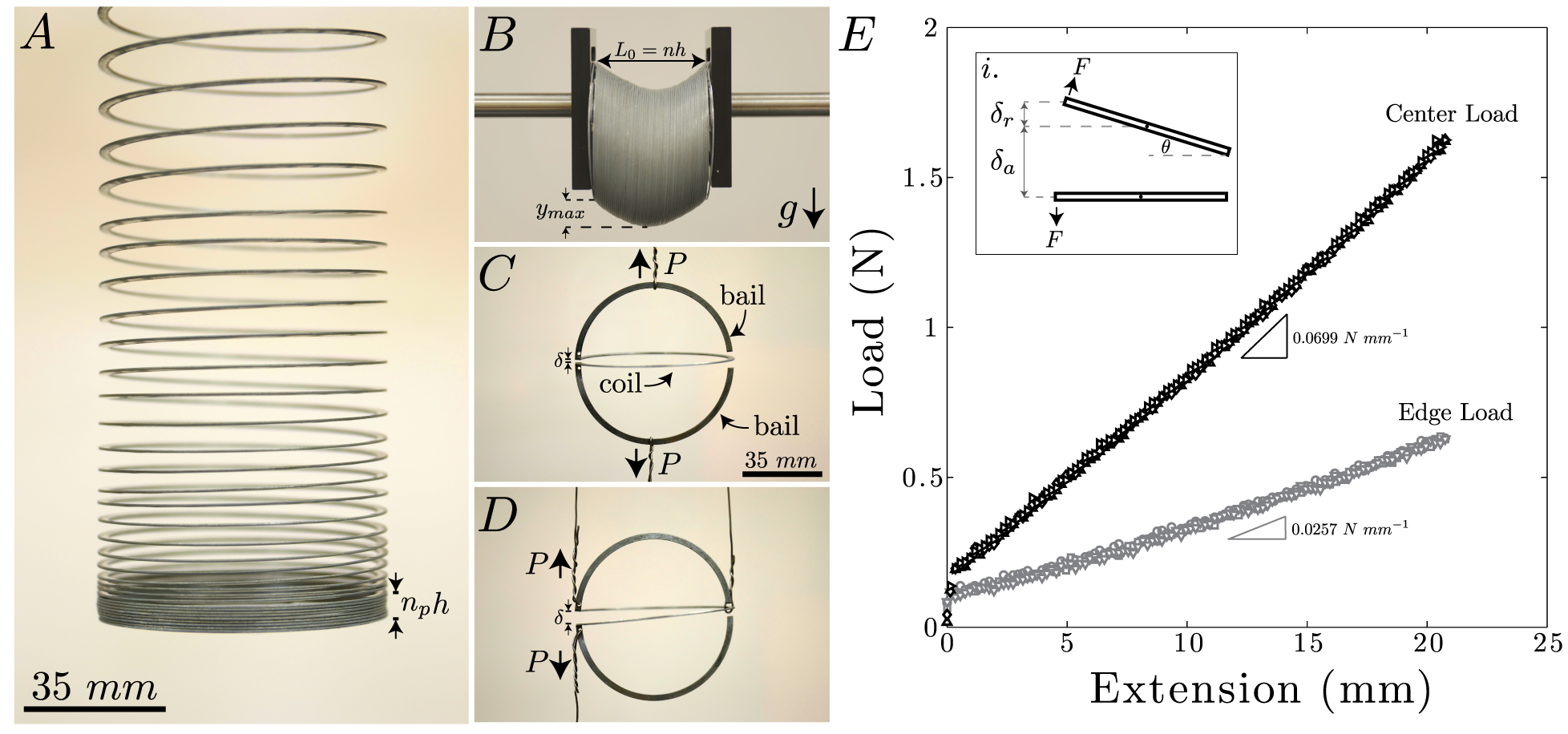}}
\end{center}
\vspace{-6mm}
\caption[]{({\bf A}) An image of the Metal (L) Slinky hanging vertically suspended at its top. ({\bf B}) An image of the same Slinky hanging horizontally in a gravitational field with its end coils held at a fixed angle of $90^{o}$, and separated by a distance $L_0$. ({\bf C}) An example of the experimental setup for the center loading and ({\bf D}) edge loading on a single coil. ({\bf E}) Force vs. displacement data for the center loaded and edge loaded coils. The slopes of these curves are used to determine $K_a$ and $K_r$, respectively. \vspace{0mm}
\label{fig-K}}
\end{figure}

The axial stiffness $K_a$ can be determined by measuring the extended length of the vertically hanging Slinky suspended at its top (Fig.~\ref{fig-K}A), and analyzing the discrete model. The compressed length of the spring is $L_0 = nh$. The extended length of the hanging model is denoted $L$ and includes the length $n_ph$ of the $n_p$ bars that are compressed together at the bottom. We define $N \equiv n-n_p$. For this vertical configuration we define the positions of the bars $y_i$ to be positive if downward, with $y_1 = 0$ at the center of the bar that is held at the top, and $L = y_{N+1} + n_ph$ where $y_{N+1}$ gives the equilibrium position of the center of the top bar among the compressed bars at the bottom. 

The governing equations are
\begin{subequations}
\begin{align}
K_a(-y_{i+1} + 2y_i - y_{i-1}) &= mg \ \text{for }  i = 2, 3,\ldots, N, \\
K_a(y_{N+1} - y_N) &= n_pmg - P_p
\end{align}
\end{subequations}
The solution is
\begin{equation}
y_i = \frac{(i - 1)}{2K_a} \left [2(n_pmg  - P_p) + (2N - i)mg\right] \ \text{for }  i = 2, 3,\ldots, N + 1 
\end{equation}
Therefore, if $P_p = n_pmg$, the extended length of the hanging Slinky model is
\begin{equation}
\label{eq-L-discrete}
L = n_ph +  \frac{N(N - 1)mg}{2K_a}
\end{equation}
Conversely, the axial spring stiffness can be obtained from equation~\ref{eq-L-discrete} as
\begin{equation}
\label{eq-Ka}
K_a =   \frac{N(N - 1)mg}{2(L-n_ph)}
\end{equation}

In the present notation, the result obtained in Mak~\cite{Mak1987} (see also~\cite{Calkin1993, Cross2012}) for a continuous spring is $K_a = N^2mg/[2(L-nh)]$. For the standard steel Slinky whose metrics are given in Table I denoted ``Metal (L)," using $n = 83$, equation~\ref{eq-Ka} results in $K_a = 64.0$ N$\cdot$ m$^{-1}$.  (Similar values were obtained by observing the lowest natural frequency of axial vibration of the hanging Slinky and comparing the measured value to the theoretical value~\cite{Heard1977}.) 

The shear stiffness can be determined by measuring the maximum deflection of the spring hanging horizontally in a gravitational field, such that the first and last coil are fixed with zero displacement, $y_1=y_n=0$. To determine the shear stiffness, the end coils are held at a fixed angle of $90^{o}$, and separated by a distance $L_0$ corresponding to the spring's compressed length (Fig.~\ref{fig-K}B). A very small initial separation beyond $L_0$ was imposed to reduce frictional effects. A force balance reveals that the shear stiffness $K_s$ to the left and right of the $i^{th}$ coil acts to resist gravity, such that $K_s(-y_{i+1}+2y_i-y_{i-1})=mg$, for $i=2,3,\ldots, n-1$. The maximum deflection depends on whether the spring contains an even or odd number of coils, with $y_{max}=n(n-2)mg(8K_s)^{-1}$ for an even number of coils, and $y_{max}=(n-1)^2(mg)(8K_s)^{-1}$ for an odd number. Therefore, the shear stiffness is given by (where $j$ denotes a positive integer)
\begin{equation}
  K_s=\begin{cases}
    \frac{mg}{8y_{max}}(n-1)^2, & \text{if } n=2j+1.\\ \\
    \frac{mg}{8y_{max}}n(n-2), & \text{if } n=2j.
  \end{cases}
\end{equation}
 For the Metal (L) Slinky in Table I, with $n=83$ and $y_{\max}=15.0$ mm, this results in a shear stiffness $K_s= 1370$ $N \cdot m^{-1}$. Table I describes the Slinkys that were tested. The symbol L denotes long, XL denotes extra long, M denotes medium length, and S denotes short. Values reported in Table 1, beyond those already described in the text, include coil thickness $h$, coil width $b$, and the mass of a single coil $m$. 

\begin{table}
\caption{\label{tab_metrics}  Slinky metrics}
\begin{ruledtabular}
\begin{tabular}{lrrrrrrrr}
   \textbf{Slinky} & \textbf{$n$} (\#) & \textbf{$L_0$}  (mm) & \textbf{$R$} (mm) & \textbf{$h$} (mm) & \textbf{$b$} (mm) & \textbf{$m$} (g) & \textbf{$\overline{EA}$} (N) & \textbf{$\overline{EI}$} (10$^{-6}$ N$\cdot$m$^2$)  \\ \hline
   Metal (L) & 82.75 & 54.82 & 34.18 & 0.67 & 2.74 & 2.49 & 0.046 & 3.15  \hspace{8mm} \\ 
   Metal (S) & 79.50 & 34.45 & 20.16 & 0.49 & 1.87 & 0.61 & 0.023 & 3.24 \hspace{8mm} \\
   Plastic (XL) & 45.50 & 148.13 & 78.50 & 2.88 & 7.40 & 14.44 & 0.099 & 582.19 \hspace{8mm} \\
   Plastic (L) & 41.00 & 77.58 & 47.47 & 1.39 & 7.77 & 3.02 & 0.019 & 47.16 \hspace{8mm} \\ 
   Plastic (M1) & 34.00 & 60.27 & 37.31 & 1.78 & 3.18 & 1.59 & 0.027 & 23.71 \hspace{8mm}  \\
   Plastic (M2) & 38.25 & 65.40 & 40.33 & 1.62 & 7.27 & 2.42 & 0.021 & 41.22 \hspace{8mm} \\
   Plastic (M3) & 37.00 & 61.42 & 38.26 & 1.66 & 3.41 & 1.65 & 0.022 & 19.41 \hspace{8mm} \\
   Plastic (S) & 31.50 & 46.97 & 31.21 & 0.93 & 6.18 & 1.20 & 0.016 & 11.27 \hspace{8mm} 
\end{tabular}
\end{ruledtabular}
\end{table}

The axial stiffness $K_a$ and rotational spring stiffness $K_r$ can also be obtained from force-displacement experiments on a single coil loaded from the center by means of bails bent outward from half coils (Fig.~\ref{fig-K}C) and the edge (Fig.~\ref{fig-K}D), respectively. The slopes of the center-loaded and edge-loaded segments in Fig.~\ref{fig-K}E are denoted $S_c$ and $S_e$, respectively. The center-loaded coil behaves like a linear spring, so that the force is simply the axial spring stiffness times the vertical displacement, and $K_a=S_c$. For the edge-loaded case, the total deflection at the edge, $\delta$, is a superposition of the axial deformation, $\delta_a$, and the bending deformation, $\delta_r$, $i.e.$ $\delta=\delta_a+\delta_r$. If the angle of splay between the coils, $\theta$, is small, then the moment about the center is $M=FR=K_r\theta\approx K_r\delta_rR^{-1}$. Therefore, we can write $\delta_a=FK_a^{-1}$ and $\delta_r = FR^2K_r^{-1}$. This leads to
\begin{equation}
\label{eq-delta}
\frac{\delta}{F} =\frac{1}{S_e}=\frac{1}{S_c}+\frac{R^2}{K_r},
\end{equation}
and hence
\begin{equation}
\label{eq-Kr}
K_r=\frac{S_cS_eR^2}{S_c-S_e}.
\end{equation}
We obtain values for the Metal (L) Slinky of $K_a=69.9 \ N \cdot m^{-1}$ and $K_r=0.047 \ N \cdot m$. The value of $K_a$ obtained from the vertically hanging Slinky is smaller than the $K_a$ obtained from force-displacement experiments by 9\%. This error may be attributed to a variation in pretension along the Slinky's length. The values obtained from the force-displacement experiments for the Metal (L) Slinky are used in the analysis below.

\section{III. Experimental Results}


\begin{wrapfigure}{r}{85mm}
\begin{center}
\vspace{-7mm}
\resizebox{0.47\textwidth}{!} {\includegraphics {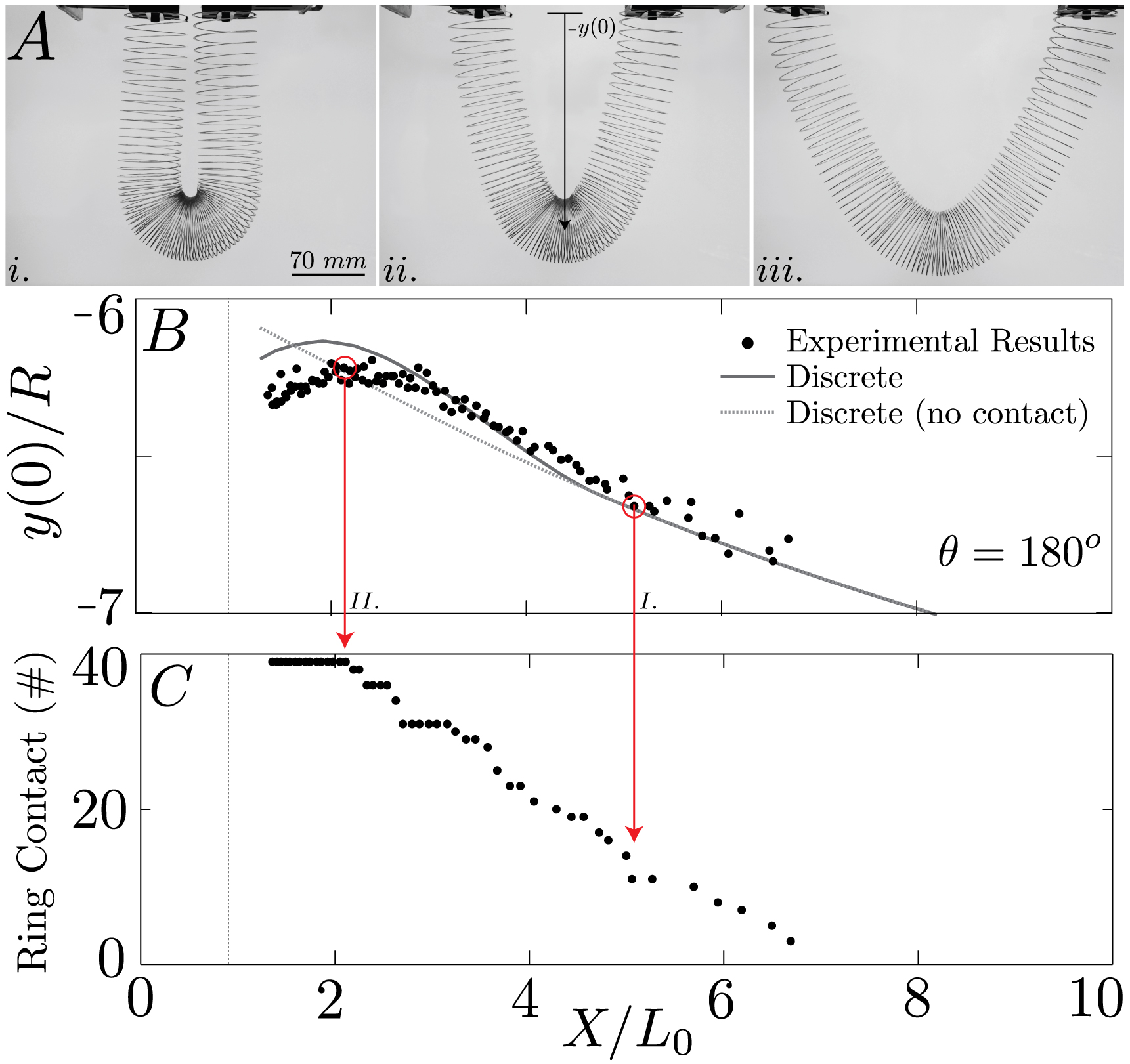}}
\end{center}
\vspace{-8.5mm}
\caption[]{({\bf A}) Images of the metal (L) Slinky held by its ends at an angle of $180^o$ and separated by $i.$ $X/L_0=2$, $ii.$ $X/L_0=5$, and $iii.$ $X/L_0=7$. ({\bf B}) A graph of the central displacement of a Slinky $y(0)$ normalized by its radius $R$ as a function of end-to-end separation $X$ normalized by the Slinky's unextended length $L_0$. Along with the experimental data, two theoretical curves are plotted - the discrete model with and without coil contact. ({\bf C}) A graph of the number of coils in contact as a function of end-to-end separation.\vspace{-3mm}
\label{fig-contact}}
\end{wrapfigure}

We first explored the various symmetric equilibrium shapes that exist when the ends of a Slinky are held at a fixed angle with $\varphi_1=\theta$ and $\varphi_n=\pi-\theta$, and their centers are separated by a finite distance (span) $X=x_n - x_1$. We measured the downward deflection $-y(0)$ of the center of the Slinky cross section at midspan as the ends were separated horizontally.  For comparison to the theoretical models presented above, the simplest configuration to consider at first is when the ends are held at $\theta=180^o$, as shown in Fig.~\ref{fig-contact}A. In this case, there is only contact between coils at the Slinky's center (if at all), and the effects of shear between coils is minimal.  In Fig.~\ref{fig-contact}B, we plot a graph of the vertical displacement of the Slinky's midpoint normalized by its radius $R$ versus the separation of the end coils normalized by the Slinky's unextended length $L_0$.  Even this fairly trivial configuration of a hanging Slinky leads to nonlinearities in its deflection as it is extended horizontally. These geometric nonlinearities emerge from both the contact between the Slinky's coils and the nonlinear terms due to the large slopes that appear in equations~\ref{eq-geo} \& b. The discrete model without consideration of contact between coils (equation~\ref{eq-V}) does not overestimate the central displacement for large values of $X/L_0$. It appears that maximal coil contact induces a significant nonlinearity in the Slinky's central displacement.  Fig.~\ref{fig-contact}C shows a corresponding graph of the number of coils in contact as a function of $X/L_0$ for the same experiment.  The contactless discrete model is able to accurately predict $y(0)/R$ when the number of coils in contact is approximately less than eleven (Fig.~\ref{fig-contact}C.).  As coil contact increases, a small degree of nonlinearity emerges in the experimental data. This nonlinearity is accurately captured when the discrete model allows for coil contact, but prevents the interpenetration of coils, as presented in equation~\ref{eq-E}. We note that additional nonlinearity is observed, and captured by our model, once the number of coils in contact reaches a fixed value, as shown by Figs.~\ref{fig-contact}B and C. 

\begin{figure}[t]
\begin{center}
\vspace{0mm}
\resizebox{1\textwidth}{!} {\includegraphics {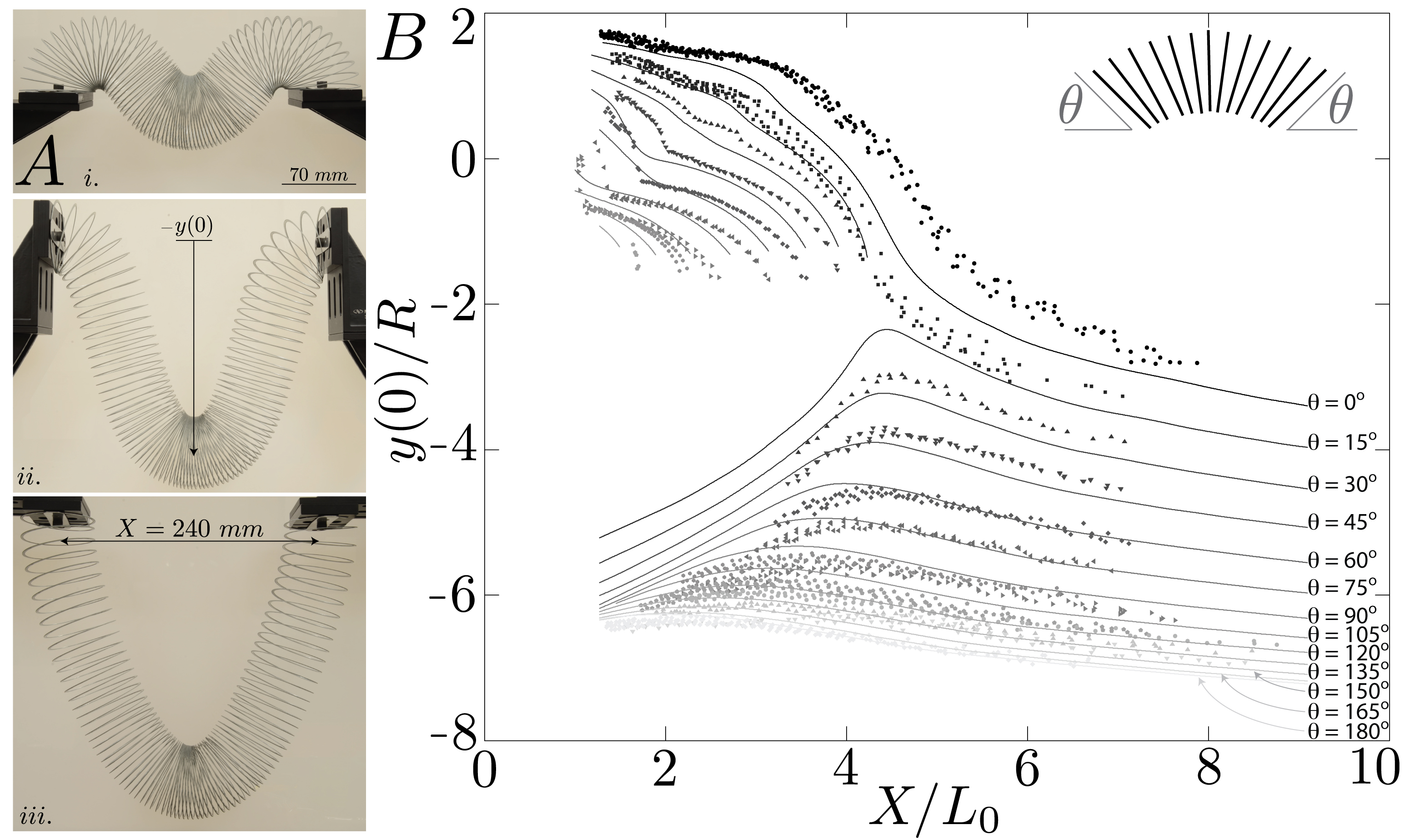}}
\end{center}
\vspace{-8mm}
\caption[]{({\bf A}) Images of the metal (L) Slinky separated by a fixed distance $X=240 \ \text{mm}$ ($X/L_0=4.38$) and held by its ends at angles of $i.$ $0^o$, $ii.$ $90^o$, and $iii.$ $180^o$. ({\bf B}) A graph of the normalized vertical midspan displacement versus normalized horizontal span of a Slinky held at angles ranging from $\theta=0^o$ to $\theta=180^o$. The solid lines correspond to numerical results from the discrete model allowing for coil contact, and the dots denote experimental results.  \vspace{-5mm}
\label{fig-horizontal}}
\end{figure}

The lateral displacement experiment was repeated for different angles $\theta$, which ranged from $\theta=0^{o}$ to $\theta=180^{o}$ in increments of $\theta=15^{o}$. Images of a horizontally extended metal (L) Slinky for three different values of $\theta$ are shown in Fig.~\ref{fig-horizontal}A. We measured the midpoint deflection as we varied the end-to-end displacement from $X/L_0=1$ to $X/L_0=9$ for each angle (Fig.~\ref{fig-horizontal}B). Three distinct deformation behaviors emerged. In the first case, which was observed for $\theta \lesssim 15^{o}$, the Slinky's arch is initially concave ({\em viz.} concave down) with its midpoint above the origin, and there is a continuous, reversible, nonlinear decrease in the Slinky's midpoint as the ends are separated horizontally. The significant geometric nonlinearities in this regime are due to both the amount of coil contact, and the distribution of this contact along the Slinky's centerline. Fig.~\ref{fig-horizontal}A-i shows coil contact at three different locations along the centerline, occurring at the midspan and the ends as well as at both the lower and upper halves of the coils. In the second case, when $30^{o} \lesssim \theta \lesssim 120^{o}$, there is a discontinuous jump in the Slinky's midpoint as it reaches a deflection of $y(0)/R \approx -2$, corresponding to an irreversible snap-through between two Slinky configurations which resembles a saddle-node bifurcation. Preceding the bifurcation, the majority of coil contact is concentrated around the Slinky's midpoint, and this nonuniform distribution of mass along the centerline is a factor in activating the snap--through. In the third case, when $\theta > 120^{o}$, the Slinky hangs with an initially convex ({\em viz.} concave up) shape, and there is very little deflection in the Slinky's midpoint as it is horizontally extended. The subtle nonlinearities in this regime were described above for the specific case of $\theta=180^o$. Theoretical predictions are plotted as solid lines along with the experimental results in Fig.~\ref{fig-horizontal}B. These curves come from minimizing the augmented total potential energy given by equation~\ref{eq-E} using the stiffness values in table 1. We note a very good qualitative agreement between our experimental and theoretical results over all displacements and edge orientations. In particular, we note that the model captures the three deformation behaviors, including the snap-through phenomenon. 


\begin{wrapfigure}{r}{82.55mm}
\begin{center}
\vspace{-8mm}
\resizebox{0.45\textwidth}{!} {\includegraphics {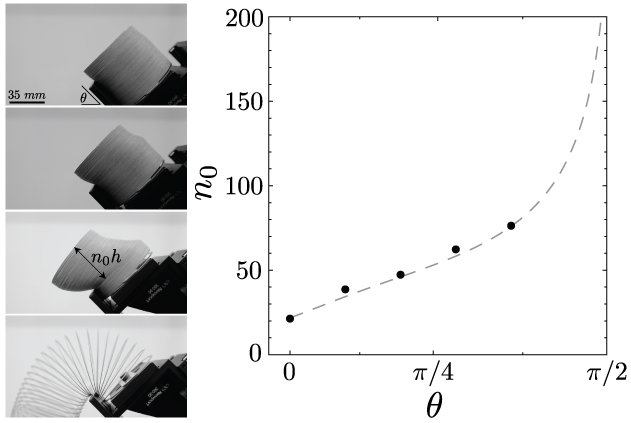}}
\end{center}
\vspace{-6mm}
\caption[]{Images of a Slinky cantilevered at an angle $\theta$. We plot $n_0$ vs. cantilever angle $\theta$. The solid curve is obtained from the discrete model (equations~\ref{eq-M1M2}a and~\ref{eq-M1M2}b), and the dots represent experimental results.  \vspace{-5mm}
\label{fig-moment}}
\end{wrapfigure}

The snap-through described above is the first example we will encounter of a large change in equilibrium shape for a small rearrangement of the Slinky's position. Multiple bifurcations between stable equilibrium shapes occur depending on the geometrical variation in the Slinky's shape. For instance, consider hanging $n_H$ coils of a Slinky upward off the edge of a surface oriented at an angle $\theta$, as shown in Fig.~\ref{fig-moment}. The pretension within the Slinky and the shearing between coils will allow this configuration to be stable up to a critical number of overhanging coils, $n_0$. The discrete model is analyzed. The stability will be determined from a balance of the moment acting on the cantilevered bars due to their weight, and the moment that resists elongation from the shear stiffness and the compressive force due to pretensioning within the Slinky. The moment at the edge of the surface is the sum of these two contributions, and stability is lost when this total moment is zero. The counterclockwise moment due to the weight of the coils is simply $M_1=mg\sum_{i=1}^{n_H}x_i$, where coil 1 is the furthest to the left, coil $n_H$ is the first overhanging one, the origin of the coordinate system is at the edge of the surface, the $x$ axis is positive to the left, and the $y$ axis is positive upward. This summation requires us to know the coordinates of the centers of mass of the overhanging bars. With $z_i$ denoting the distance (positive if upward) along overhanging bar $i$ from a leftward 	extension of the surface (at angle $\theta$ with the $x$ axis) to the bar's center of mass, equilibrium along bar $i$ yields $K_s(-z_{i-1} + 2z_i - z_{i+1}) = -mg \cos \theta$ for $i = 2,3,\ldots, n_H$ where $z_{n_H+1} = R$, and $K_s(z_1 - z_2) = -mg \cos \theta$. Then, from geometry, one can show that the locations of the centers of mass of the overhanging bars are
\begin{subequations}\label{eq-moment-xyz}
\begin{align}
z_i&=R-\frac{mg \cos \theta}{2K_s}(n_H+i)(n_H+1-i) \\
x_i&=\frac{h \cos \theta}{2}(1+2n_H-2i)-z_i \sin \theta \\
y_i&=\frac{h \sin \theta}{2}(1+2n_H-2i)+z_i\cos \theta 
\end{align}
\end{subequations}
Since the pretension $P_p=mgn_p$ acts through the center of bar $n_H$, we can write the competing moment as $M_2=-mgn_pz_{n_H}$, positive if counterclockwise about the edge.  Using equations~\ref{eq-moment-xyz}a--c, we find that
\begin{subequations}
\label{eq-M1M2}
\begin{align}
M_1&=mg\left[-n_HR\sin \theta+\frac{h}{2}n_H^2\cos\theta+\frac{mg}{6K_s}n_H\left(2n_H^2+3n_H+1\right)\sin\theta\cos\theta\right] \\
M_2&=-n_pmg(R-\frac{mg}{K_s}n_H\cos\theta)
\end{align}
\end{subequations}
The critical number of cantilevered coils is found by setting $M_1+M_2=0$, which leads to a cubic equation for $n_H$. The closest integer greater than the lowest real solution $n_H$ yields the critical value $n_0$, and failure is expected (see lowest photograph in Fig.~\ref{fig-moment}) if $n_0$ Slinky coils overhang the edge, according to the discrete model. In Fig.~\ref{fig-moment}, we show a cantilevered Slinky, along with a plot of the critical number of cantilevered coils $n_0$ as a function of angle $\theta$.   The equation for the critical number of cantilevered coils is plotted in Fig.~\ref{fig-moment} for the Metal (L) Slinky, $i.e.$ $m=0.00249 \ \text{kg}$, $h=0.00067 \ \text{m}$, $n_p=5$, and $R=0.03418 \ \text{m}$. This Slinky has a shear stiffness of $K_s=1320 \ \text{N m}^{-1}$.  There is good agreement between our model and experimental results denoted by dots.

\begin{wrapfigure}{r}{82.55mm}
\begin{center}
\vspace{0mm}
\resizebox{0.45\textwidth}{!} {\includegraphics {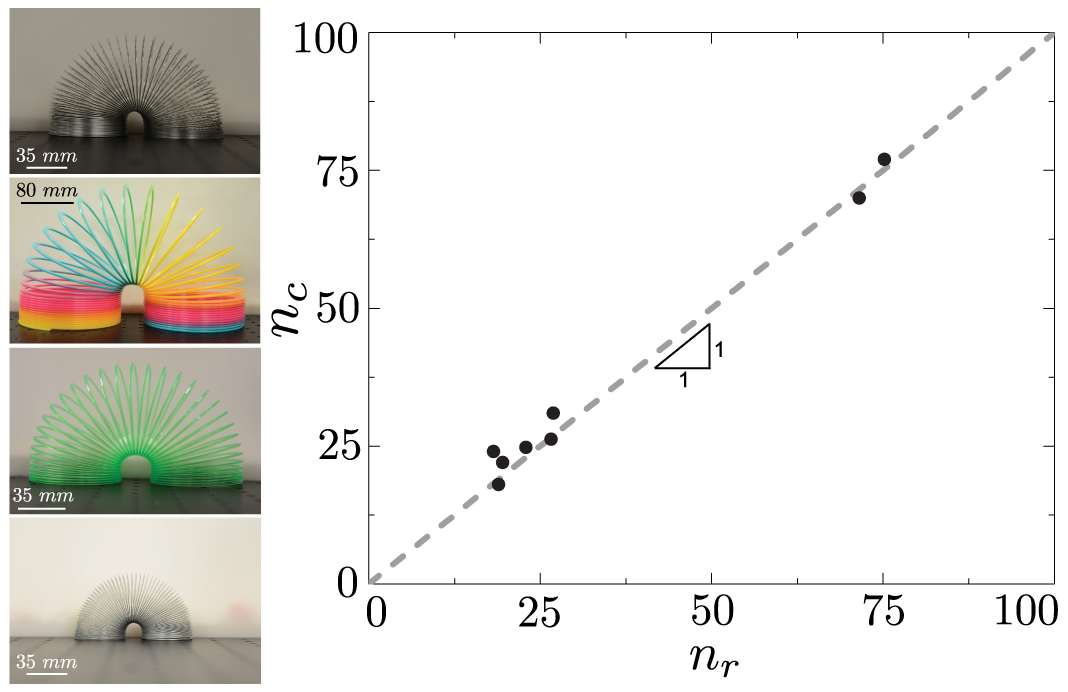}}
\end{center}
\vspace{-6mm}
\caption[]{Images of various Slinkys in an arch shape. When the number of coils in the Slinky is below a critical value $n_c$, it is no longer able to form an arch. We plot this critical parameter vs. $n_r$ from equation~\ref{eq-slinkyness}, with dots denoting experimental results.  \vspace{0mm}
\label{fig-slinkyness}}
\end{wrapfigure}

With a strong correlation between our model using the Slinky's mechanical properties, and the equilibrium shapes of the Slinky, we can generalize this model to a spring of any material or size by nondimensionalizing the relevant parameters.  We normalize the total effective energy as $\overline{V}=\frac{2V}{nmgR}$, and the axial deformation $\Delta \xi_i$ and vertical displacement $y_i$ by the coil thickness $h$, such that $\Delta \overline{\xi}_i=\Delta \xi_i/h$ and $\overline{y}_i=y_i/h$.  Due to the large separation of scales between shear and either bending or axial deformation, we neglect the shear stiffness and pretension, and write the dimensionless form of equation~\ref{eq-V} as
\begin{equation}
\label{eq-V-nondim}
\overline{V}=\frac{\overline{EA}h}{nmgR}\sum_{i=1}^{n-1}\Delta \overline{\xi}_i^2+ \frac{\overline{EI}}{nmgRh}\sum_{i=1}^{n-1}\Delta \varphi_i^2+\frac{2h}{nR}\sum_{i=1}^{n}\overline{y}_i,
\end{equation}
where the barred quantities $\overline{EA}$ and $\overline{EI}$ represent effective axial and bending stiffnesses of the helical spring, respectively. These quantities are directly related to spring stiffnesses described in section II, with $\overline{EA}=K_ah$ and $\overline{EI}=K_rh$. Equation~\ref{eq-V-nondim} provides several nondimensional quantities that we can use to describe the various stability criteria of the Slinky. For instance, the prefactor to the first summation in equation~\ref{eq-V-nondim} represents a balance between axial extension and gravity, $i.e.$ a spring with $n \sim \overline{EA}h/mgR$ will extend beyond $L_0$ if held vertically from its top in a gravitational field. The second summation represents a balance between bending stiffness and gravity, which provides a scaling of the number of coils in a spring required for the structure to bend into a stable arch, 
\begin{equation}
\label{eq-slinkyness}
n_r\sim \frac{\overline{EI}}{mgRh}
\end{equation}
We tested the validity of this scaling on a variety of flexible springs that were initially stable as both arches and cylinders. Individual coils, or fractions of coils, were removed until the Slinky was unable to form a stable arch. We note that between the arch and the cylinder configurations, a stable, intermediate state occurs in which one arch base rotates and only contacts the surface at a point. We measured the critical number of coils $n_c$ required to form a stable arch with both bases in axial contact with a horizontal surface ($\theta=0$) for a variety of commercially available flexible springs (Fig.~\ref{fig-slinkyness}). Values of $\overline{EA}$ for each Slinky were obtained as described in section II, while $\overline{EI}$ values were obtained using Castigliano's method~\cite{Borum2014}. We plot $n_c$ versus $n_r$, given by equation~\ref{eq-slinkyness}, as the horizontal axis. The dots denote experimental results corresponding to the Slinky examples listed in Table 1, and the dashed line represents $n_c=n_r$. The scaling in equation~\ref{eq-slinkyness} is in good agreement with the experimental results. 

\begin{wrapfigure}{r}{82.55mm}
\begin{center}
\vspace{-8mm}
\resizebox{0.45\textwidth}{!} {\includegraphics {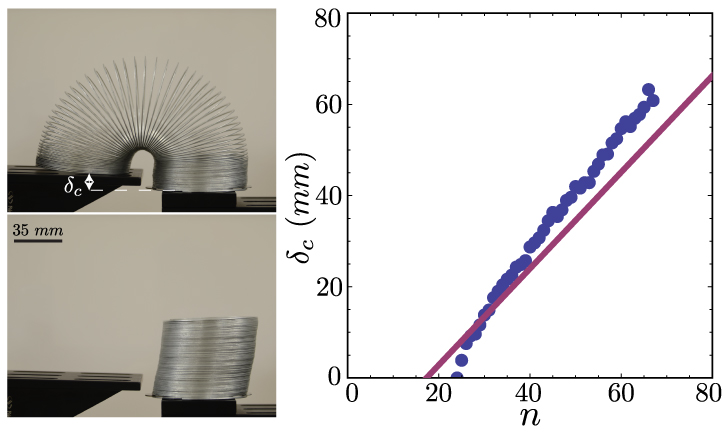}}
\end{center}
\vspace{-6mm}
\caption[]{Images of Slinky losing stability as one edge is lowered below a critical displacement $\delta_c$, and a plot of $\delta_c$ vs. the number of coils $n$.  \vspace{-8mm}
\label{fig-tilt}}
\end{wrapfigure}

Once a Slinky is stable in the shape of an arch, stability loss can occur if one end of the spring is lifted above a critical height, which we refer to as the step instability.  Experimentally, we incrementally decreased $y_n$ relative to $y_1$ in a quasi-static manner (where the $y$ axis is upward), and measured the critical displacement $\delta_c=y_1-y_n$ as a function of the number of coils $n$ (Fig.~\ref{fig-tilt}). This vertical displacement instability is similar to the one described above for the number of coils required to stably form an arch. Decreasing the magnitude of $y_n$ by a height equivalent to a coil's thickness, $i.e.$ $\delta=h$, relative to $y_1$ is analogous to removing a single coil from the Slinky. Therefore, the effective number of coils in the Slinky is simply $n_{eff}=n-\delta/h$. This effective coil number is similar to the scaling in equation~\ref{eq-slinkyness}, however there will be axial resistance as one end of the Slinky is lowered in addition to the Slinky's rotational stiffness. By observing that all the coils in Fig.~\ref{fig-tilt} are in contact, we note that the Slinky satisfies the constraint described by equation~\ref{eq-xi}. If we neglect shear and assume that $\Delta \varphi_i$ for all $i$ are small, we have
\begin{equation}
\label{eq-xi-approx}
\Delta \xi_i = 2R \sin\frac{\Delta \varphi_i}{2}+h\cos\frac{\Delta \varphi_i}{2} \approx R \Delta \varphi_i + h
\end{equation}
This approximation allows the nondimensional potential energy given in equation~\ref{eq-V-nondim}, leaving out terms that are constant or are linear in $\Delta \varphi_i$, to be rewritten as
\begin{equation}
\label{eq-V-nondim-tilt}
\overline{V}=\frac{\overline{EA}h}{nmgR}\sum_{i=1}^{n-1}\left(\frac{R\Delta \varphi_i}{h}\right)^2+ \frac{\overline{EI}}{nmgRh}\sum_{i=1}^{n-1}\Delta \varphi_i^2+\frac{2h}{nR}\sum_{i=1}^{n}\overline{y}_i =\left(\frac{\overline{EA}R^2+\overline{EI}}{nmgRh}\right)\sum_{i=1}^{n-1}\Delta \varphi_i^2+\frac{2h}{nR}\sum_{i=1}^{n}\overline{y}_i.
\end{equation}

The prefactor of the first summation on the right hand side of the equation essentially describes the dimensionless balance between axial and rotational stiffness and gravity when there is contact between all the coils, 
\begin{equation}
\label{eq-nar}
n_{ar}\sim\frac{\overline{EA}R^2+\overline{EI}}{mgRh}
\end{equation}
We set the effective coil number $n_{eff}$ equal to $n_{ar}$ to solve for the critical vertical displacement, and obtain:
\begin{equation}
\label{eq-dc}
\delta_c\approx nh-\frac{\overline{EA}R^2+\overline{EI}}{mgR}
\end{equation}

We see in Fig.~\ref{fig-tilt} that, for the Metal (L) Slinky, equation~\ref{eq-dc} captures the general trend of the data (denoted by dots). The discrepancy with the data is likely due to contributions from shear and pretension which are neglected in the scaling presented in equation~\ref{eq-nar}.

\vspace{-3mm}

\section{Conclusions}
In this work, we present a discrete model to capture a Slinky's static equilibria and unstable transitions. The model considers the Slinky's axial, shear, and rotational stiffnesses, and calculates the equilibrium shapes that result from a minimization of the structure's total potential energy augmented by penalty functions to account for coil contact. We emphasize that modeling the contact between coils is crucial for describing its equilibrium shapes and quasi-static stability criteria.  We determined the flexible spring's stiffnesses by isolating specific static equilibrium shapes.  Finally, we provide a general description of highly flexible helical springs by considering the nondimensional potential energy of the spring, enabling the formulation of parameters that may describe and explain a Slinky's stability behavior under a variety of actions. The focus of this work was on configurations for which the locus of the centers of the coils is planar. Relaxing this planar configuration would be a natural extension of the current work. 

\section{Acknowledgements}
The authors acknowledge Poof--Slinky, Inc. for donating the initial Slinkys used in this work, and Virginia Tech's Department of Engineering Science and Mechanics for use of shared facilities. The work of A.D. Borum was supported by the NSF-GRFP under Grant No. DGE-1144245.

\bibliographystyle{unsrt}

\end{document}